\documentstyle[seceq,epsf,wrapft,twoside]{ptptex}
\notypesetlogo  %comment in if to eliminate PTPTeX logo
\title{Properties of $\sigma$ and $\kappa$ Production Amplitudes 
}
\author{%
Muneyuki {\sc Ishida} and Shin Ishida$^*$
}
\inst{%
Department of Physics, Meisei University, 
Hino 191-8506, Japan\\
$^*$  Research Institute of Quantum Science, 
College of Science and Technology\\
Nihon University, Tokyo 101-0062, Japan
}
\abst{%
Our method of analysis, which led to existence of $\sigma$ and $\kappa$ mesons, is
reviewed and examined from a viewpoint of general $S$-matrix. 
It is shown that the method is consistent with the constraints from chiral symmetry and unitarity.
Accordingly the long-believed common analyses of $\pi\pi (K\pi )$ scattering and production processes,
based on elastic unitarity, prove to lose its theoretical base.
The observed phase motion by 180 degrees of $\sigma$ shows also the validity of our method.
}
\begin{document}
\maketitle

\setcounter{tocdepth}{4}

\section{Introduction}

The  existence of the iso-singlet scalar $\sigma$ meson 
had been a longstanding problem of hadron physics.
Conventional analyses\cite{pen} 
of $\pi\pi$ scattering phase shift\cite{CM} lead to a conclusion of no $\sigma$ existence, 
while recent reanalyses\cite{555} including ours suggest the existence of 
light $\sigma (450$-$600)$.
The result of the previous analysis with no $\sigma$ existence 
was pointed out\cite{frascati} to be not correct, since in this analysis\cite{pen}
there is no consideration on the cancellation mechanism between 
$\sigma$ amplitude and non-resonant $\pi\pi$ amplitude, which is guaranteed by chiral symmetry.  

It is remarkable that, in contrast with the spectra of $\pi\pi$ scattering, 
the clear peak structure has been observed in mass region of $m_{\pi\pi}\sim 500$ MeV 
in the various $\pi\pi$ production processes, such as 
$J/\psi \rightarrow \omega\pi\pi$,\cite{DM2,had97Jpsi,WuNing,WNII}
$p\bar p\rightarrow 3\pi^0$,\cite{CB,ppbar}
$D^+\rightarrow \pi^+\pi^-\pi^-$\cite{E791} and 
$\tau^-\rightarrow \pi^-\pi^0\pi^0 \nu_\tau$,\cite{tau} and this structure 
is shown to be well reproduced by the Breit-Wigner amplitude of $\sigma$ meson.
Presently firm evidences\cite{YITP} of $\sigma$ seem to be accumulated, 
and the column of $\sigma$ in particle data group table is corrected as 
``$f_0(600)$ or $\sigma$" in the newest'02\cite{PDG} edition 
in place of $f_0(400$--1200) or $\sigma$ in the '96--'00 editions.

There are now hot controversies on the existence of $I=1/2$ scalar 
$\kappa$ meson, to be assigned as a member of $\sigma$ nonet.
Reanalyses\cite{kappa,Kpi} of $K\pi$ scattering phase shift\cite{LASS} 
suggest existence of the $\kappa (900)$, while no $\kappa$ is insisted in ref. \citen{penkappa}.
The existence of $\kappa$ is again suggested strongly
in $K\pi$ production process of $D^+\rightarrow K^-\pi^+\pi^+$\cite{E791kappa}
and $J/\psi \rightarrow \bar K^{*0}K^+\pi^-$\cite{WuNing,WNII,Komada},
similarly to the case of $\sigma$.

In the analyses of the $\pi\pi$ (or $K\pi$) production processes mentioned above,
the amplitudes are parametrized by a coherent sum of the Breit-Wigner amplitudes 
including $\sigma$ (or $\kappa$) and of the non-resonant $\pi\pi$ (or $K\pi$) production amplitude. 
This parametrization method is called VMW method.\cite{Sawazaki,ShinMune} 
However, this method was strongly criticized\cite{MO1,bug} from the conventional viewpoint, 
so called ``universality argument."\cite{pen}
In this argument it is insisted that all the $\pi\pi$($K\pi$) production amplitude ${\cal F}$
and the scattering amplitude ${\cal T}$ have a common phase 
due to the Watson final state interaction theorem (being based upon the elastic unitarity), 
and both the $\pi\pi$($K\pi$) production and scattering processes must be analyzed 
together with the common phase of the amplitudes.

In this talk we shall review our method of analysis and examine the
the relation between ${\cal T}$ and ${\cal F}$
from a viewpoint of generalized  $S$ matrix.
As a result,
we emphasize that the ${\cal F}$ is, in principle, independent from the ${\cal T}$ and 
that the analyses of production processes should be done independently of scattering processes.
Accordingly our method of analyses, VMW method, along this line of thought is consistent\cite{PLB2} 
with all the constraints from unitarity and chiral symmetry. 
The experimental phase motion of the production amplitude is also examined, and 
the above criticisms on VMW method will be shown not to be correct also experimentally.

\section{Properties of $\pi\pi /K\pi$ Scattering amplitude}

We first review our reanalysis\cite{555} of $\delta_S^0$, $\pi\pi$ scattering phase shift 
of $I=0$ $S$ wave amplitude,
obtained by CERN-Munich.\cite{CM} The applied method is 
Interfering Amplitude method, where the total $\delta_S^0$ below 
$m_{\pi\pi}\simeq 1$GeV is represented by the sum of the component phase shifts,
\begin{eqnarray}
\delta_S^0 &=& \delta_\sigma + \delta_{BG} + \delta_{f_0} .
\label{b1}
\end{eqnarray}
The $\delta_\sigma$ and $\delta_{f_0}$ are, respectively, contributions from  
$\sigma$ and $f_0(980)$ Breit-Wigner amplitudes.
The $\delta_{BG}$ is from non-resonant repulsive $\pi\pi$ amplitude, 
which is taken phenomenologically of hard-core type, 
$\delta_{BG}=-p_1 r_c$ ($p_1=\sqrt{s/4-m_\pi^2}$ being the CM momentum of $\pi$).

The experimental $\delta_S^0$ passes through 90$^\circ$ at $\sqrt s(=m_{\pi\pi})\sim 900$MeV. 
This is explained by the cancellation between attractive $\delta_\sigma$ and repulsive 
$\delta_{BG}$. (See ref. \citen{555}.) 
The mass and width of $\sigma$ is obtained as $m_\sigma =585\pm 20$MeV and
$\Gamma_\sigma =385\pm 70$MeV.

Note that the above cancellation is shown\cite{frascati} to come from chiral symmetry
in the linear $\sigma$ model (L$\sigma$M): The $\pi\pi$ scattering $A(s,t,u)$ amplitude 
in L$\sigma$M is given by 
\begin{eqnarray}
A(s,t,u) &=& \frac{(-2g_{\sigma\pi\pi})^2}{m_\sigma^2-s} - 2 \lambda 
 = \frac{s-m_\pi^2}{f_\pi^2} + \frac{1}{f_\pi^2} 
\frac{(s-m_\pi^2)^2}{m_\sigma^2-s},\ \ \ \ \ 
\label{b2}
\end{eqnarray}
as a sum of the $\sigma$ amplitude $A_\sigma$, which is strongly attractive, and of the 
non-resonant $\pi\pi$ amplitude $A_{\pi\pi}$ due to the $\lambda\phi^4$ interaction, which 
is stongly repulsive. They cancel with each other following the relation of L$\sigma$M,
 $g_{\sigma\pi\pi}=f_\pi\lambda=(m_\sigma^2-m_\pi^2)/(2f_\pi)$,
and the small ${\cal O}(p^2)$ Tomozawa-Weinberg (TW) amplitude and its correction are left.
The $A_\sigma (A_{\pi\pi})$ corresponds to $\delta_\sigma$($\delta_{BG}$).
Actually, the theoretical predictions for $\delta_{NR}^0$ and $\delta_{NR}^2$, obtained by 
unitarizing $A_{\pi\pi}$, $A(t,s,u)$ and $A(u,t,s)$ in L$\sigma$M, are consistent with our
$\delta_{BG}$ of hard core type in our phase shift analysis,\cite{555}
and with experimental $\delta_0^2$,\cite{hoog} respectively (See ref. \citen{frascati,SUNY}). 

Thus, it is shown that the $\sigma$ Breit-Wigner amplitude with non-derivative 
(${\cal O}(p^0)$) $\pi\pi$-coupling requires at the same time the strong 
(${\cal O}(p^0)$) repulsive $\pi\pi$ interaction to obtain the small ${\cal O}(p^2)$ 
TW amplitude, satisfying chiral symmetry. This is the origin of
$\delta_{BG}$ in our phase shift analysis.\footnote{
There was an argument\cite{MO1} that a broad resonance with mass 1GeV,
denoted as $f_0(1000)$\cite{CM} or $\epsilon (900)$,\cite{pen} instead of light $\sigma$, exists.
However, in this argument the above cancellation mechanism is not considered, and their conclusion
of no existence of light $\sigma$ is not correct.
}

The similar cancellation is also expected to occur in $K\pi$ scattering since
$K$ has also a property of Nambu-Goldstone boson. The experimental $I=1/2$ $S$ wave phase shift
$\delta_{S}^{1/2}$ passes through 70 degrees at about $m_{K\pi}\sim 1.3$GeV. This $\delta_{S}^{1/2}$ 
is parametrized by introducing the $\kappa$ Breit-Wigner phase shift $\delta_\kappa$ and its 
compensating repulsive non-resonant $K\pi$ phase shift $\delta_{BG}^{Non.Res}$ as 
$\delta_{S}^{1/2}=\delta_\kappa + \delta_{BG}^{Non.Res}+\delta_{K_0^*(1430)}$.
The fit\cite{Kpi} to $\delta_{S}^{1/2}$ below 1.6 GeV by LASS\cite{LASS} gives the mass and width of 
$\kappa$ meson\cite{Kpi} as $m_\kappa =905\stackrel{+65}{\scriptstyle -30}$ MeV
and $\Gamma_\kappa =545\stackrel{+235}{\scriptstyle -110}$ MeV.

As we explained above, because of the chiral cancellation mechanism,
the $\pi\pi /K\pi$ scattering amplitude ${\cal T}$ has the spectra strongly suppressed 
in the threshold region. Correspondingly ${\cal T}$ shows slowly varying phase motion 
in the $\sigma /\kappa$ mass region(, totally about 90 degrees being much smaller than 180 degrees
given by the $\sigma (\kappa )$ Breit-Wigner amplitude itself).
We consider whether these two features of  ${\cal T}$ are also valid in general  
$\pi\pi /K\pi$ production amplitudes ${\cal F}$ or not, in the following sections.  

\section{Method of Analyses of $\pi\pi /K\pi$ Production Processes}

\hspace*{-0.8cm}($Essence$ $of$ $VMW$ $method$)\ \ \ \ 
The analyses of $\pi\pi /K\pi$ production processes 
quoted in \S 1 are done following the VMW method.\cite{Sawazaki,ShinMune}
Here we explain our basic physical picture on strong interactions and the essential
point of this method.

The strong interaction is a residual interaction of QCD among all 
color-neutral bound states of quarks($q$), anti-quarks($\bar q$) and gluons($g$). 
These states are denoted as $\phi_i$, and the strong interaction 
Hamiltonian ${\cal H}_{\rm str}$ is described by these 
$\phi_i$ fields.
It should be noted that, from the quark physical picture,\cite{ShinMune} 
unstable particles
as well as stable particles, if they are color-singlet bound states, 
should be equally treated as $\phi_i$-fields on the same footing.
\begin{eqnarray}
{\cal H}_{\rm str} &=& {\cal H}_{\rm str}  (\phi_i) \nonumber\\
\{\ \phi_i \ \} &=& 
 \{  {\rm color\ singlet\ bound\ states\ of\ }q,\bar q\ {\rm and}\ g   \}    .
\label{c1}
\end{eqnarray}
%where the fields with underline are unstable particles. The  
%various resonant particles are included in the set of $\phi_i$'s.

The time-evolution by ${\cal H}_{\rm str} (\phi_i)$ describes 
the generalized $S$-matrix. 
Here, it is to be noted that, if  
 ${\cal H}_{\rm str}$ is hermitian, the unitarity of $S$ matrix is guaranteed.
 
The bases of generalized $S$-matrix 
are the configuration space of these multi-$\phi_i$ states. 
\begin{eqnarray}
\begin{array}{l} S\ {\rm matrix\ bases}\\ 
 \ \{ {\rm multi-}\phi_i{\rm -states} \} \end{array}
 &=& 
   \left\{  \begin{array}{l}   
  | \omega \pi\pi \rangle , \underline{ | \omega\sigma \rangle , 
                   | \omega f_2 \rangle ,| b_1 \pi \rangle , \cdots , |J/\psi\rangle }, \\
  | N\pi \rangle ,|N\pi\pi\rangle ,
                 \underline{ |\Delta\rangle ,|\Delta\pi\rangle ,
         |\Delta\sigma\rangle ,\cdots},\\ 
  \cdots\cdots 
   \end{array}     \right\}  ,\ \ \ \ \ \ \ \ \ \ 
\label{c2}
\end{eqnarray}
where the states relevant for $J/\psi\rightarrow\omega\pi\pi$ decay
and $N\pi$ scattering are respectively shown in 1st and 2nd lines as examples.
The states including unstable particles shown with underlines are
equally treated with non-resonant states $|\omega\pi\pi\rangle$ and $|N\pi\rangle,|N\pi\pi\rangle$.
%%%%%%%%%%%%%%%%%%%%%%%%%%%

\begin{figure}
  \epsfysize=2.5cm
   \centerline{\epsffile{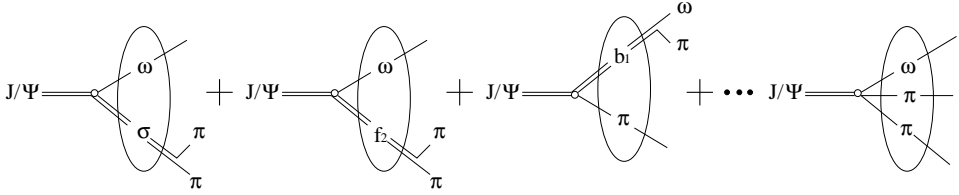}}
\caption{The final state interactions in $J/\psi\rightarrow\omega\pi\pi$.
The amplitude is a superposition of different $S$ matrix elements, such as
$J/\psi\rightarrow\omega\sigma$,  $J/\psi\rightarrow\omega f_2$,  
$J/\psi\rightarrow b_1 \pi$, $\cdots$,  $J/\psi\rightarrow\omega (\pi\pi)_{Non.Res.}$.
The ellipses represent the final state interactions, 
and the corresponding amplitudes have independent strong phases.
}
 \label{figFSI}
\end{figure}

The relevant $J/\psi \rightarrow \omega\pi\pi$ decay process 
is described by a coherent sum of different non-diagonal 
elements of 
%the generalized $S$-matrix (\ref{c2.2}):
various 2-body decay amplitudes, 
${}_{out}\langle \omega\sigma | J/\psi\rangle_{in}$, ${}_{out}\langle \omega f_2 | J/\psi\rangle_{in}$,
 ${}_{out}\langle b_1 \pi | J/\psi\rangle_{in}$, $\cdots$, and a non-resonant 
3-body($\omega\pi\pi$) decay amplitude${}_{out}\langle \omega\pi\pi | J/\psi\rangle_{in}$ .
They have mutually independent coupling strengths, $r_{\psi\sigma}$, $r_{\psi f_2}$, $r_{\psi b_1}$, 
$\cdots$, $r_{\psi\pi\pi}$.\footnote{
The strengths and phases of respective amplitudes are considered to be determined 
by quark dynamics. However, we treat them independent in phenomenological analyses.
}
The ${\cal H}_{\rm str}$ induces the various final state interaction, 
reducing to the strong phases of the corresponding amplitudes, 
$\theta_{\psi\sigma}$, $\theta_{\psi f_2}$, $\cdots$ .
(See Fig.\ref{figFSI}.) These phases are related with the phases of diagonal elements
$\theta_{\psi}$, $\theta_{\sigma}$, $\cdots$ which are unknown.
Thus, we may treat phenomenologically the formers as independent parameters from the latters.    

The remaining problem is how to treat unstable particles, as
there is no established field-theoretical method for this problem. 
In VMW method the decay of the unstable particles are treated intuitively 
by the replacement of propagator,
\begin{eqnarray}
{\rm Stable\ particle} & & {\rm Unstable\ particle}\nonumber\\
\frac{1}{m_\sigma^2-s-i\epsilon} 
&\stackrel{\rm Strong\ Int.}{\longrightarrow}&
\Delta_\sigma (s)=\frac{m_\sigma \Gamma_\sigma}{m_\sigma^2-s-im_\sigma\Gamma_\sigma (s)} ,
\label{c3}
\end{eqnarray}
where we take the case of $\sigma$ as an example.
Here we should note that
the imaginary part of the denominator $i m_\sigma \Gamma_\sigma (s)$ does not come from the 
$\pi\pi$ final state interaction (or the repetition of the virtual $\pi\pi$ loops), but from the decay of $\sigma$ to 
$\pi\pi$ state.\footnote{
In non-relativistic quantum mechanics, the decay of the unstable particle can be described 
by the WF with imaginary part in the energy, $e^{-iE_0t}\rightarrow e^{-i(E_0-i\Gamma /2)t}$. Correspondingly, the propagator is replaced as 
$\frac{1}{E_0-E}\rightarrow \frac{1}{E_0 - E - i \Gamma /2}$ .
Similarly, in the Feynman propagator, 
$\Delta_F(x)=\int \frac{d^4k}{(2\pi )^4}\frac{e^{ikx}}{m^2+k^2-i\epsilon}
=\int \frac{d^3{\bf k}}{i(2\pi )^32\omega}e^{-i\omega |x_0|}e^{i{\bf k}\cdot{\bf x}} $, when we replace 
 $e^{-i\omega |x_0|}$ by  $e^{-i (\omega -i\Gamma /2)|x_0|}$, we obtain the propagator
$\frac{1-i\Gamma /(2\omega )}{m^2+k^2-\Gamma^2/4-i\omega\Gamma}$, which may be approximated as
$\frac{1}{m^2-s-i m \Gamma (s)}$ in Eq.~(\ref{c3}).
}

As a result, 
the effective $\omega\pi\pi$ amplitude is given by 
%a coherent sum of all these decay amplitudes,
\begin{eqnarray}
  {\cal F}_{ \omega  \pi  \pi} \ \ \  &=& 
   {\cal F}_{\omega\sigma}+ {\cal F}_{\omega f_2}  
   +{\cal F}_{ b_1\pi} +\cdots + {\cal F}_{\omega (\pi\pi)_{Non.Res.}} ,
\label{c4}\\
 {\cal F}_{\omega\sigma}\ \ \ 
     &=& {}_{out}\langle \omega\sigma | J/\psi \rangle_{in}\Delta_\sigma (s) 
     = r_{\psi\sigma} e^{i\theta_{\psi\sigma}}  
       \frac{m_\sigma \Gamma_\sigma}{m_\sigma^2-s-im_\sigma\Gamma_\sigma (s)} \nonumber\\
{\cal F}_{\omega f_2}\ \ \  
     &=& {}_{out}\langle \omega f_2 |\  J/\psi\  \rangle_{in} \Delta_{f_2}(s)
     = r_{\psi f_2} e^{i\theta_{\psi f_2}} 
       \frac{m_{f_2} \Gamma_{f_2} N_{\pi\pi}(s,{\rm cos}\theta)}
              {m_{f_2}^2-s-im_{f_2}\Gamma_{f_2} (s)} \nonumber\\
{\cal F}_{\omega b_1}\ \ \  
     &=& {}_{out}\langle \omega b_1 |J/\psi \rangle_{in} \Delta_{b_1}(s) 
     =r_{\psi b_1} e^{i\theta_{\psi b_1}} 
        \frac{m_{b_1} \Gamma_{b_1}}
              {m_{b_1}^2-s-im_{b_1}\Gamma_{b_1} (s)}  \nonumber\\
     \cdots &&  \nonumber\\
  {\cal F}_{\omega (\pi\pi)}^{Non.Res.} 
     &=& {}_{out}\langle \omega (2\pi)_{N.R.} | J/\psi \rangle_{in} \ \ \ \ \ 
     =\ \ \ r_{2\pi}^{N.R.} e^{i\theta_{2\pi}^{N.R.}}\ \ , 
\label{c5}
\end{eqnarray}
where ${\cal F}_{\omega\sigma}$, ${\cal F}_{\omega f_2}$, $\cdots$ 
corresponds to the generalized $S$ matrix 
element $\ {}_{out}\langle \omega\sigma |\  J/\psi\  \rangle_{in}$,
$\ {}_{out}\langle \omega f_2 |\  J/\psi\  \rangle_{in}$, respectively.
(The extra Breit-Wigner factor comes from the prescription Eq.~(\ref{c3}), and
 $N_{\pi\pi}$ is angular function of $f_2\rightarrow \pi\pi$ $D$-wave decay.)
They have mutually-independent couplings, $r_{\psi\sigma}$, $r_{\psi f_2}$, $r_{\psi b_1}$, $\cdots$ 
and $r_{\psi2\pi}^{N.R.}$. The strong phases $\theta_{\psi\sigma}$, $\theta_{f_2}$, etc.
are  treated as free parameters, as we explained previously.
The amplitude given by Eqs.~(\ref{c4}) and (\ref{c5}) is consistent with the $S$-matrix unitarity(generalized unitarity).
This parametrization method is called as VMW method.

\hspace*{-0.5cm}({\it Chiral constraint and threshold suppression})\ \ \ \ 
In $\pi\pi$ scattering the derivative-coupling 
property of Nambu-Goldstone $\pi$-meson requires 
the suppression of the amplitude
${\cal T}_{\pi\pi}$ near threshold,
\begin{eqnarray}
{\cal T}_{\pi\pi} &\sim& -p_{\pi 1}\cdot p_{\pi 2} \rightarrow  m_\pi^2 \sim 0 \ \ \ {\rm at} \ \ \ 
s\rightarrow 4m_\pi^2.
\label{c6}
\end{eqnarray}
This chiral constraint requires, as was explained in \S 2, the
strong cancellation between the $\sigma$ amplitude ${\cal T}_\sigma$ and the
non-resonant $\pi\pi$ amplitude ${\cal T}_{2\pi}$, which 
means the strong constraints, $r_{2\pi}\simeq -r_\sigma$, 
$\theta_\sigma \simeq \theta_{2\pi}$, in the corresponding formulas to 
Eqs.~(\ref{c4}) and (\ref{c5}). 
The amplitude has zero close to the threshold and 
no direct $\sigma$ Breit-Wigner peak is observed in $\pi\pi$ mass spectra. 

On the other hand, for general $\pi\pi$ production processes 
the parameters $r_i$ and $\theta_i$ are independent of those in  
$\pi\pi$ scattering, since they are concerned with different 
$S$-matrix elements.
Especially we can expect in the case of ``$\sigma$-dominance", 
$r_\sigma \gg r_{2\pi}$, the $\pi\pi$ spectra 
show steep increase from the $\pi\pi$ threshold, and the 
$\sigma$ Breit-Wigner peak is directly observed. 
This situation seems to be realized in 
$J/\psi\rightarrow\omega\pi\pi$ and $D^+\rightarrow\pi^-\pi^+\pi^+$.

Here we should note that the chiral constraint on 
$r_\sigma$ and $r_{2\pi}$ does not work 
generally in the production processes 
with large energy release to the $\pi\pi$ system.
We explain this fact in case of $\Upsilon$ decays.\cite{PLB2} Here we take 
$J/\psi\rightarrow\omega\pi\pi$ as an example.
We consider a non-resonant $\pi\pi$ amplitude of derivative-type 
${\cal F}_{\rm der}$, 
\begin{eqnarray}
{\cal F}_{\rm der} &\sim&  P_\psi \cdot p_{\pi 1} P_\psi \cdot p_{\pi 2} / M_\psi^2 , 
\label{c7}
\end{eqnarray}
where $P_\psi (p_{\pi i})$ is the momentum of $J/\psi$ 
(emitted pions).\footnote{
The equation (\ref{c7}) is obtained by the chiral symmetric effective 
Lagrangian,
${\cal L}_d=\xi_d \partial_\mu \partial_\nu \psi_\lambda \omega_\lambda 
(\partial_\mu \pi \partial_\nu \pi + \partial_\mu \sigma \partial_\nu \sigma )$.
The possible origin of this effective Lagrangian is discussed in ref.\cite{PLB2}. 
There occurs no one $\sigma$-production amplitude, cancelling the $2\pi$ amplitude,  
in this Lagrangian. 
}
This type of amplitude satisfies the Adler self-consistency condition,
${\cal F}\rightarrow 0$ when $p_{\pi 1\mu}\rightarrow 0$, and consistent with
the general chiral constraint. 
However, this zero does not appear in low energy region of actual $s$-plane.
At $\pi\pi$ threshold (where $s=4m_\pi^2$),
$p_{\pi 1\mu}= p_{\pi 2\mu}$ and  
$P_\psi \cdot p_{\pi i} / M_\psi=E_{\pi i} =
(M_\psi^2 -m_\omega^2 + 4 m_\pi^2 )/(4M_\psi )$ ($E_{\pi i}$ being the energy of emitted pion), 
and thus, 
\begin{eqnarray}
{\cal F}_{\rm der} &\rightarrow& \{\ (M_\psi -(m_\omega^2 - 4 m_\pi^2)/M_\psi )/4\ \}^2 \gg m_\pi^2 \ \ \ {\rm at}\ \ \ 
s\rightarrow 4m_\pi^2.
\label{c8}
\end{eqnarray}
The amplitude (\ref{c7}) is not suppressed near $\pi\pi$ threshold, and 
correspondingly there is no strong constraint between $r_{\psi\sigma} e^{i\theta_{\psi\sigma}}$
and $r_{2\pi}e^{i\theta_{2\pi}}$ leading to the threshold suppression.
This is quite in contrast with the situation in $\pi\pi$ scattering, 
Eq.~(\ref{c6}). 

\hspace*{-0.5cm}($``Universality"$ $of$ ${\cal T}_{\pi\pi}$ : $threshold$ $behavior$)\ \ \ \ 
Conventionally all the production amplitudes ${\cal F}_{\pi\pi}$,
including the $\pi\pi$ system in the final channel, 
are believed\cite{pen} to take the form 
proportional to ${\cal T}_{\pi\pi}$ as
\begin{eqnarray}
{\cal F}_{\pi\pi} &=& \alpha (s) {\cal T}_{\pi\pi};
\ \ \alpha (s):{\rm slowly\ varying \ 
                       real\ function} , 
\label{c9}
\end{eqnarray} 
where $\alpha (s)$ is supposed to be a slowly varying real function.
This implies that 
${\cal F}$ and ${\cal T}$ have the same phases and the same structures
(the common positions of poles, if they exist).
The equation (\ref{c9}) is actually applied to the analyses of various production  
processes\cite{pen,bug,Zou}, and it was the reason of 
overlooking $\sigma$
for almost 20 years in the 1976 through 1994 editions of Particle Data Group tables. 

The equation (\ref{c9}) is based on the belief that 
low energy $\pi\pi$ chiral dynamics
is also applicable to $\pi\pi$ production processes with small $s$,
leading to the threshold suppression of spectra, as in  Eq.~(\ref{c6}), 
in all production processes
because of the Adler zero in ${\cal T}_{\pi\pi}$. This 
is apparently inconsistent with experimental data.

So, in order to remove the Adler zero at $s=s_0$ and 
to fit the experimental spectra, one is forced to modify\cite{pen,bug} 
the form of $\alpha (s)$ by multiplying artificially 
the rapidly varying factor $1/(s-s_0)$ without any theoretical reason.

Here I should like to stress the physical meaning of an example, Eq.~(\ref{c7}),
that Adler zero condition, even in the isolated final $2\pi$ system, does not necessarily
lead to the threshold suppression. 
This implies that the above mentioned ad hoc prescription\cite{bug}
to get rid of the undesirable zero near threshold of ${\cal F}$ 
becomes not necessary, if we take the new dynamics, which would appear with large energy
release, duely into account.

\section{Phases of Production Amplitudes}

($Generalized$ $S$ $matrix$ $and$ $phase$ $of$ $production$ $amplitude$)\ \ \ \ 
It is often discussed that in order to confirm the existence of a resonant 
particle, it is necessary to observe the corresponding phase motion $\Delta \delta \sim 180^\circ$ 
of the amplitude. In the $\pi\pi\ P$ wave amplitude 
a clear phase motion $\Delta \delta \sim 180^\circ$ due to $\rho$ meson Breit-Wigner amplitude 
is observed. However, in the case of $\sigma$ meson, because of the chiral cancellation 
mechanism (explained in \S 2), and of its large width, its 
phase motion cannot be observed directly in the $\pi\pi$ 
scattering amplitude. Because, {\it in production processes, the amplitudes are a sum of 
various $S$ matrix elements
(as explained in \S 3), 
the pure $\sigma$ Breit-Wigner phase motion may be generally difficult 
to be observed.} 
However, in some exceptional cases, when the amplitude is dominated by $\sigma$,
it may be directly observed. 

On the other hand,  
it is widely believed that all the $\pi\pi$ production amplitudes ${\cal F}$
have the same phase as that of the scattering amplitude.
%%%%%%%%%
In the relevant $J/\psi \rightarrow \omega\pi\pi$ 
decay,
it is often argued that the ${\cal F}$ near $\pi\pi$ threshold 
must take the same phase as the scattering phase $\delta$, 
since in this energy region the $m_{\omega\pi} (\sim M_\psi )$ is large
and $\pi\pi$ decouples from $\omega$ in the final channel.\cite{Tuan} 

However, {\it this belief (or Eq.~(\ref{c9})) comes from an 
improper application of the elastic unitarity condition,
which is not applicable to the production processes},
where the freedom of various strong phases, 
$\theta_{\psi\sigma}$, $\theta_{\psi b_1}$, $\theta_{\psi f_2}$,
$\cdots$  
is allowed from the {\it generalized unitarity condition}.

{\it Because of the above strong phases, generally ${\cal F}$ 
have different phases from ${\cal T}$.}  
The ${\cal F}$ has the same phase as ${\cal T}$ only
when the final $\pi\pi$ 
systems are isolated in strong interaction level. For 3-body decays such as 
$J/\psi\rightarrow \omega\pi\pi
% (K^*K\pi)
$ and $D^-\rightarrow\pi^-\pi^+\pi^+
%(K^-\pi^+\pi^+)
$, the above condition is not satisfied.
Actually, the large strong phases in $J/\psi$ and $D$ decays are suggested
experimentally.
In order to reproduce the branching ratios of 
$J/\psi\rightarrow 1^-0^-$ decays 
(that is, $J/\psi\rightarrow \omega\pi^0,\rho\pi,K^*\bar K,\cdots$,)
it is necessary to introduce a large relative strong phase\cite{Mahiko} 
$\delta_\gamma =arg\frac{a_\gamma}{a}=80.3^\circ$ 
between the effective coupling constants
of three gluon decay $a$ and of one photon decay $a_\gamma$.
A similar result is also obtained in  $J/\psi\rightarrow 0^-0^-$ decays.
%($J/\psi\rightarrow\pi^+\pi^-,\ K^+K^-$ and $K^0\bar K^0$).
A large relative phase between $I=3/2$ and $I=1/2$ amplitudes 
of $D\rightarrow K\pi$ decays is observed: 
$\delta_{3/2}(m_D)-\delta_{1/2}(m_D)=(96\pm 13)^\circ$,\cite{Dphase}
(while in $B\rightarrow D\pi ,D\rho ,D^*\pi$ decays rather small relative phases are obtained).
By considering these results we expect existence of not small strong phases 
$\theta_{\psi\sigma}$, $\theta_{\psi b_1}$,
$\cdots$ coming from $\sigma \omega ,b_1\pi ,\cdots$ rescatterings 
in $J/\psi$ decays.

%%%%%%%%%%%%%%%%%%%%%%%%%%%
According to these works\cite{Mahiko,Dphase} $M_\psi$  ($M_D$) are not sufficiently large 
for making $\pi\pi$ decouple from $\omega$ ($\pi_1^-\pi_2^+$ from $\pi_3^+$).
The $\pi\pi$ elastic unitarity is not valid
in the amplitude ${\cal F}$ Eq.~(\ref{c4}), which takes the different phase from ${\cal T}$.
Similar consideration is also applicable to the $J/\psi\rightarrow K^*K\pi$ and 
$D^-\rightarrow K^-\pi^+\pi^+$, and the corresponding $K\pi$ production amplitudes ${\cal F}$
take different phases from the phase of $K\pi$ scattering amplitude ${\cal T}$.
%%%%%%%%%%%%%%%%%%%%%%%%%%%

(180$^\circ$ $phase$ $motion$ $of$ $\sigma$-$meson$ $observed$ $in$ 
$D^+\rightarrow \pi^-\pi^+\pi^+$)\ \ \ \ 
Recently a method extracting the $\sigma$ phase motion from Dalitz plot data of 
$D^+\rightarrow\pi^-_1\pi^+_2\pi^+_3$ decays
is presented in ref. \cite{brazil}, where 
the interference between the $f_2(1275)$ Breit-Wigner amplitude in 
$s_{12}=m_{\pi^-_1\pi^+_2}^2\simeq (1.275$GeV$)^2$ region 
and the remaining $S$-wave component in $s_{13}=m_{\pi^-_1\pi^+_3}\simeq (0.5$GeV$)^2$ region is used. 
The actual analysis of the E791 data gives
the clear 180 degrees phase motion in $s_{13}\simeq 0.25$GeV$^2$ region which is 
consistent with that\cite{Gobel} of the $\sigma$ Breit-Wigner amplitude.
This phase motion is completely different from the $\pi\pi$ scattering phase shift,
where the phase moves only 90 degrees below 900 MeV. 
Thus, {\it it is experimentally confirmed that the phase of $\pi\pi$ production amplitude 
${\cal F}$ in this process 
is different from that in $\pi\pi$ scattering amplitude ${\cal T}$} and that
{\it the $\pi\pi$ elastic unitarity does not work in $D$ decays.} 
The $180$ degrees phase motion seems to suggest experimentally 
the $\sigma$ is not the state from final $\pi\pi$ interaction
but corresponds to the real entity in quark level forming the general $S$-matrix bases.   

Similar result is also obtained for $\kappa$ in $D^+\rightarrow K^-\pi^+\pi^+$.
The large phase motion of $K\pi$ $S$ wave component around $\kappa$ energy region is suggested, 
which is much larger than the $K\pi$ scattering phase shift obtained by LASS experiment.\cite{LASS}

\section{Concluding Remarks}

For many years it had been believed that both the scattering and the production $\pi\pi /K\pi$
amplitudes in the low mass region with $\sqrt s < 1$GeV have the same structures 
(universality of scattering amplitudes), and that analyses of any production process should be done together
with the scattering process. Because of this belief the peak structures (due to $\sigma /\kappa$ meson)
in the various production processes had been regarded as mere backgrounds.
However, in this talk we have explained that the above belief is not correct and that the production processes
should generally be treated independently from the scattering process.
The essential reason of this(, as is explained in \S 3,) is that as the basic fields of expanding the $S$ matrix,
the ``bare" fields of $\sigma$ and $\kappa$ (as the bound states of quarks), 
as well as the $\pi$ and $K$ fields, should be taken into account.

%\acknowledgements
%%
%%The author would like to express his sincere gratitudes
%%to the organizing committee, especially to Prof. A. Fariborz and Prof. J. Schechter,
%%for giving him an opportunity of presenting a talk. 
%
%The authors express their thanks to Prof. S. F. Tuan for useful information.
%They are also grateful to Prof. M. Oka, Prof. K. Takamatsu and 
%Prof. T. Tsuru for discussions, and to Dr. T. Ishida for making figures.


\begin{thebibliography}{99}
\bibitem{pen} K. L. Au, D. Morgan and M. R. Pennington, Phys. Rev. D {\bf 35} (1987), 1633.
%\\
% D. Morgan and M. R. Pennington, Phys. Rev. D {\bf 48} (1993), 1185.
\bibitem{CM} B. Hyams et al., Nucl. Phys. B {\bf 64} (1973), 134.
% W. Ochs, Ludwig-Maximilians-University Munich, thesis 1973.\\
G. Grayer, Nucl. Phys. B {\bf 75} (1974), 189.
\bibitem{555} E. Beveren et al., Z. Phys. {\bf C30} (1986), 651.\\
N. N. Achasov and G. N. Schestakov, Phys. Rev. D {\bf 49} (1994), 5779.\\
R. Kaminski, L. Lesniak and J.-P. Maillet, Phys. Rev. D {\bf 50} (1994), 3145.\\
S. Ishida, M. Ishida, H. Takahashi, T. Ishida, K. Takamatsu and
T. Tsuru, Prog. Theor. Phys. {\bf 95} (1996), 745; {\bf 98} (1997), 1005.\\
N. A. Tornqvist and M. Roos, Phys. Rev. Lett. {\bf 76} (1996), 1575.\\ 
M. Harada, F. Sannino and J. Schechter, Phys. Rev. D {\bf 54} (1996), 1991.\\
A. Dobado and J. R. Pelaez, Phys. Rev. D {\bf 56} (1997), 3057.\\
J. A. Oller, E. Oset and J. R. Pelaez, Phys. Rev. D {\bf 59} (1999) 074001; Erratum D {\bf 60} (1999) 099906.\\
G. Colangelo, J. Gasser and H. Leutwyler, Nucl. Phys. B {\bf 603} (2001), 125. 
\bibitem{frascati} M. Ishida, Prog. Theor. Phys. {\bf 96} (1996), 853;
proc. of ``Workshop on Hadron Spectroscopy"(WHS99), 
Frascati, Frascati Physics series 15 (1999), hep-ph/9905261.
\bibitem{DM2} J. E. Augustin et al.(DM2 collaboration), Nucl. Phys. B {\bf 320} (1989), 1. 
\bibitem{had97Jpsi} K. Takamatsu et al., proc. of 7th int. conf. on ``Hadron Spectroscopy",
Hadron'97 at BNL, AIP conf proc. 432. hep-ph/9712232. 
\bibitem{WuNing} Wu Ning (BES collaboration), hep-ex/0104050.
\bibitem{WNII} Wu Ning, hep-ex/0304001; in proc of ``Hadron Spectroscopy, Chiral Symmetry and 
Relativisitic Description of Bound Systems," Nihon univ and KEK symposium, 
held at Ichigaya, Tokyo, February 24-26, 2003.
\bibitem{CB} C. Amsler et al.(Crystal Barrel collaboration), Phys. Lett. B {\bf 342} (1995), 433.
\bibitem{ppbar} M. Ishida et al.,
%, T. Komada, S. Ishida, T. Ishida, K. Takamatsu and T. Tsuru,
Prog. Theor. Phys. {\bf 104} (2000), 203.
\bibitem{E791} E. M. Aitala et al. (E791 collaboration), Phys. Rev. Lett. {\bf 86} (2001), 770;765.
\bibitem{tau} D. M. Asner et al. (CLEO collaboration), Phys. Rev. D {\bf 61} (2000), 012002.
\bibitem{YITP} N. A. Tornqvist, in proc. of ``Possible Existence of $\sigma$ Meson
and Its Implication to Hadron Physics"($\sigma$ Meson 2000), YITP Kyoto, June 2000,
Soryushiron kenkyu (Kyoto) {\bf 102} No. 5 (2001); KEK-proceedings 2000-4.
\bibitem{PDG} K. Hagiwara et al, Phys. Rev. {\bf D66} (2002), 010001.
\bibitem{kappa} 
E. Beveren et al., Z. Phys. {\bf C30} (1986), 651.\\
D.~Black, A.~H.~Fariborz, F.~Sannino and J.~Schechter, Phys. Rev. {\bf D58} (1998), 054012.
\bibitem{Kpi} S. Ishida et al., Prog. Theor. Phys. {\bf 98} (1997), 621.
\bibitem{LASS} D. Aston et al. (LASS collaboration), Nucl. Phys. B {\bf 296} (1988), 493.
\bibitem{penkappa} S. N. Cherry and M. R. Pennington, Nucl. Phys. A {\bf 688} (2001), 823.
\bibitem{E791kappa} E.M.Aitala et al.(E791 collaboration), 
Phys.~Rev.~Lett.~{\bf 16} (2003),~121801;~hep-ex/0204018.
%C. Gobel, hep-ex/0110052.
\bibitem{Komada} T. Komada in proc of Nihon univ and KEK symposium.
\bibitem{Sawazaki} S. Ishida, M. Oda, H. Sawazaki and K. Yamada, Prog. Theor. Phys. {\bf 82} (1989), 119.\\
S. Ishida, in proc of Hadron97, BNL 1997, AIP conf proc 432, Upton NY ed. by S. U. Chung and H. J. Willutzki; 
in proc. of WHS99. 
\bibitem{ShinMune} S. Ishida and M. Ishida, proc. of $\sigma$ Meson 2000, YITP Kyoto. kep-ph/0012323.
\bibitem{MO1} P. Minkowski and W. Ochs, hep-ph/0209225; Eur. Phys. J. C {\bf 9} (1999), 283.
\bibitem{bug} D. V. Bugg, private communication.
\bibitem{PLB2} M. Ishida, S. Ishida, T. Komada and S. I. Matsumoto, 
       Phys. Lett. {\bf B518} (2001), 47.
\bibitem{hoog} W. Hoogland et al., Nucl. Phys. B {\bf 126} (1977), 109; B {\bf 69} (1974), 266.
\bibitem{SUNY} M. Ishida, in proc of ``Scalar meson workshop," held at SUNY (Utica), May 2003.
%\bibitem{MO2} P. Minkowski and W. Ochs, Eur. Phys. J. C {\bf 9} (1999), 283.
%\bibitem{PLB1} T. Komada, M. Ishida and S. Ishida, Phys. Lett. B {\bf 508} (2001), 31.
\bibitem{Zou} B. S. Zou and D. V. Bugg, Phys. Rev. D {\bf 48} (1993), R3948; D {\bf 50} (1994), 591.
\bibitem{Mahiko} M. Suzuki, Phys. Rev. D {\bf 58} (1998), 111504. 
hep-ph/9801284; 9801327; 0001170.
\bibitem{Dphase} M. Bishai et al., (CLEO collaboration), Phys. Rev. Lett. {\bf 78} (1997), 3261. 
\bibitem{Tuan} Informed from S. F. Tuan. This condition is suggested by V. V. Anisovich.\\
S. F. Tuan, summary talk in Nihon univ. and KEK symposium.
\bibitem{brazil} I. Bediaga and J. M. de Miranda, hep-ph/0211078; Phys. Lett. {\bf B550} (2002), 135.
\bibitem{Gobel} C. Gobel, in proc of Nihon uinv and KEK symposium; talk in this workshop.

\end{thebibliography}
\end{document}